\documentstyle[12pt,aasms4]{article}

\slugcomment{revised, re-submitted version, v5 8/98}

\begin{document}

\title { THE STELLAR CONTENT OF 10 DWARF IRREGULAR GALAXIES } 
\author{REGINA E. SCHULTE-LADBECK} 
\affil{University of Pittsburgh, Department of Physics and
Astronomy, 3941 O'Hara St., Pittsburgh, PA 15260, U.S.A.} 
\authoremail {rsl@phyast.pitt.edu}

\author{ULRICH HOPP}
\affil{Universit\"{a}tssternwarte M\"{u}nchen, Scheiner Str.~1, D-81679
M\"{u}nchen, F.R.G.}
\authoremail {hopp@usm.uni-muenchen.de}

\begin{abstract}

We examine the stellar content of 10 dwarf irregular galaxies of which 
broad-band CCD photometry was published in Hopp \& Schulte-Ladbeck (1995). 
We also present H$_{\alpha}$ images for several of these galaxies. 
The galaxies in the sample are located outside of the Local Group.
Yet, they are still close enough to be resolved into single stars from 
the ground but only the brightest stars (or star clusters) are detected 
and there is severe crowding. The sample galaxies were selected to be 
isolated from massive neighbors; about half of them are (mostly peripheral) 
members of groups, the other half is located in the field. We discuss the
vicinity of the sample galaxies to other dwarf galaxies.

In order to interpret single-star photometry and draw conclusions about
the stellar content or other distance-dependent quantities, it is crucial
that accurate distances to the galaxies be known. The distances to the sample 
galaxies are not well known since all but one have not had a primary
distance indicator measured. We make an attempt to constrain the distances by
identifying the envelope of the brightest supergiants in B, B-R and R, B-R
color-magnitude diagrams, but the results are not very accurate (we estimate
the minimal error on the distance modulus is 1.36$^m$). Nevertheless, 
the fact that the sample galaxies are resolved with direct ground-based imaging 
indicates that they are sufficiently nearby to represent
good candidates for observations with instruments that provide high spatial 
resolution, e.g., adaptive optics systems on large ground-based telescopes, 
or the Hubble Space Telescope.

We discuss the morphologies, color-magnitude diagrams and frequencies
of the resolved stars together with the morphology of the ionized gas, as well as 
the surface brightness profiles and colors of the underlying light 
distributions of unresolved stars. We point out the occurrence in half 
of the galaxies studied of HII regions and young stellar associations located well
outside of the main body of resolved stars. This appears to be in conflict
with the hypothesis of self-propagating star-formation. All of the sample 
galaxies contain HII regions and young massive stars with ages of a few Myr
to around 10~Myr. For supergiants beyond an age of about 50~Myr, incompleteness
is already a problem in the single-star photometry. However, we can also gain 
insight into the stellar content from the integrated colors of the unresolved stars. 
The light distribution of the unresolved stars is more extended
than that of the resolved stars and is of a more regular and elliptical
shape. We provide ellipticities, central surface brightnesses and scale lengths
for the sample galaxies. The background-light colors indicate a range of star-formation
histories for the sample galaxies, with galaxy colors at one extreme
being dominated by the old, metal-poor population, and at the other extreme,
by the most recent star-birth event. The results provide insight into the stellar 
content and the star-formation histories of isolated, late-type galaxies.

\end{abstract}

\keywords{galaxies: Irregular}

\section{INTRODUCTION}

Irregular galaxies (IGs) represent a class of low-mass galaxies with
active star formation. 
The basic, integrated properties of IGs have been summarized by 
Gallagher \& Hunter (1984) and Hunter \& Gallagher (1986). 
IGs can be found isolated in the field and without any near and massive 
neighbors. Such isolated IGs that can be well resolved into single stars 
are excellent laboratories to study the star-formation and
propagation processes intrinsic to IGs. Star formation can apparently 
be a self-initiating process in isolated IGs, since 
it is unlikely to have been triggered by galaxy-galaxy interactions, a mode
of star formation suggested to apply to many starburst galaxies (Larson \&
Tinsley 1978). While Brinks (1990) has proposed that interactions of dwarf
irregular galaxies (dIs) with other dwarf galaxies could stimulate star 
formation in dIs, and Taylor et al. (1995, 1996) have found an appreciable 
number of 
HI companions to several star-forming dIs, it still remains unclear whether 
interactions are a trigger of current star formation even for those dIs 
which have companions (Skillman 1994). 

IGs are structurally simple objects which rotate more slowly than 
spirals (Gallagher \& Hunter 1984)
and so star formation is also unlikely to be induced by density waves. 
While some IGs are presently undergoing global bursts of 
star formation (see Gallagher \& Hunter 1984 and references therein), 
star formation in most IGs has been suggested to be a local phenomenon on
theoretical grounds, starting in one part of the galaxy
and then self-propagating to other parts, as the supernovae originating 
from the evolution of massive stars in one star-forming region 
induce star formation in adjacent regions of the galaxy (Gerola 
\& Seiden 1978).

A census of the resolved, luminous stellar content of nearby IGs can be taken 
with photometry from the ground. Local IGs thus furnish nearby and 
resolved analogs of somewhat more distant HII galaxies, which, in most 
cases, have not been resolved into single stars and of which only 
integrated spectra can be analyzed 
(but see, e.g., Meurer et al. 1994, Schulte-Ladbeck et al. 1998). 
Nearby IGs may also represent present-epoch examples for the large
numbers of galaxies with irregular morphologies discovered at redshifts 
of 0.3 to 0.5 in the Hubble Space Telescope (HST)
Medium Deep Survey. The evolution of IGs may explain the ``faint 
blue galaxy'' problem, the excess numbers of blue galaxies observed in
deep images of the sky (e.g., Griffiths et al. 1996, 
Glazebrook et al. 1995). Studying isolated IGs locally will allow us to
better understand the morphology and stellar content of such 
galaxies and will have bearing on the question of whether the 
faint blue galaxies can be explained with intrinsically star-forming 
dwarfs or whether they are merging progenitors of today's galaxies. 
Isolated, nearby IGs can therefore play an important role for our 
understanding of the stellar populations of distant galaxies.

Owing to the small resolutions and limiting magnitudes achievable from the 
ground, most studies of the resolved stars in IGs have concentrated on the 
Local Group (cf. Table~1 in Bresolin et al. 1993, discussion in Greggio
1995, Gallart et al. 1996, Table~2, Grebel 1997). However, it cannot be 
excluded that the star-formation history of Local Group dwarfs has been
influenced by their proximity to the Galaxy and M31 (e.g., van den 
Bergh 1994). Thus, if we want to gain insight into the intrinsic 
star-forming properties of IGs, we have to look beyond the Local Group.
Our work has centered on galaxies which are located beyond the Local Group, 
but which are still close enough to be resolved into single stars
from the ground (Hopp \& Schulte-Ladbeck 1987, 1991, 1995; 
Schulte-Ladbeck \& Hopp 1995). While several studies of IGs have achieved
deeper limiting magnitudes than our's (see above, also Tolstoy 1995),
all of these studies focussed on the exploration of a single, or very
few, galaxies. The comparison of the single-star photometry of the
IGs studied by different groups has been notoriously difficult
owing to the different filters and different filter systems used by
the different investigators, different reduction methods, etc.
Thus, previous comparisons have been hampered by uncertainties 
in the magnitude conversions between different filter systems, and
have been limited to a few galaxies. We here present a homogeneous
dataset of single-star photometry for 10 galaxies. 

Our study, however, has its own limitations. Due to the large distances of the sample
galaxies only the brightest and hence, most luminous objects are resolved. They
may be star clusters rather than single stars. Even in such nearby objects
as the LMC, resolving clusters of massive stars into single objects has
presented difficulty (e.g., the case of R136, see e.g., Weigelt et al. 1991). Any
single-star photometry of the fainter, older populations (like asymptotic
giant branch stars, red giant stars) will have to await
high-resolution imaging. Data of sufficient quality to resolve old
populations can be obtained with the HST
for galaxies even beyond the Local Group (e.g., Schulte-Ladbeck
et al. 1998). In this paper, only the integrated light of the
faint stars is analyzed. Therefore, this paper is only a first step towards 
characterizing the star-forming properties of distant
dIs from the investigation of their resolved stellar content.

\section{THE SAMPLE}

The sample was drawn from the volume-limited catalog of nearby galaxies by 
Kran-Korteweg (1986, KK). The galaxies in our sample were selected because 
of their relative isolation from massive galaxies. In Table 1, we give an 
overview of the sample; all data are from KK. The first column is the UGC 
number of the galaxy, the second column provides other names, the third 
column is the galaxy type, the fourth column gives the group membership 
classification, the fifth and sixth column provide the size of the major 
axis in arcsec and the ellipticity
(${\epsilon}$=1-b/a), the seventh and eighth columns are the apparent 
blue magnitude, B$_T$, and the apparent blue magnitude corrected for 
galactic and internal absorption, B$_T$$^0{}^,{}^i$, the ninth
and tenth column list the distance in units Mpc for a Virgo cluster infall 
model with v$_V{}_C$=220~km~s$^-{}^1$, and the corresponding 
absolute blue magnitude based on this model, M$_B$$^0{}^,{}^i$, and 
corrected for galactic and internal extinction, while the eleventh and 
twelfth column provide the 
same values for the model with v$_V{}_C$=440~km~s$^-{}^1$. The Virgo Cluster 
distance used by KK is 21.7 Mpc. Notice that UGC~8091=GR~8 is classed
``Field, LG?" in KK. Its location is certainly at the fringe of the
Local Group. Hodge (1974), e.g., placed it within the Local Group.
Recently, a single Cepheid was discovered in GR~8 by Tolstoy et al. (1995),
who suggest that this dI is more distant than previously thought. 

Considering now the location of the galaxies in our sample with respect 
to the spatial distribution of other nearby galaxies, as well as the numbers 
and distances of their nearest neighbors, we may better recognize the 
relative ``isolation" of the sample galaxies and rank-order them by 
increasing isolation 
or decreasing environmental galaxy density. Among the galaxies located 
in groups (see Table 1) UGC~7559 is the only galaxy found within the 
inner, denser region of its group. UGC~4459, UGC~8024, UGC~8091, and UGC~8320
are situated in the outskirts of their respective groups and hence, 
in low-density regions. While UGC~8320 is listed as a field galaxy
in KK, we caution that there are four relatively nearby galaxies.
UGC~5272~A,B, UGC~5340, and UGC~8760 are all in the field and 
have few, distant neighbors; UGC~6456 is the most isolated of the galaxies 
in our sample.

The term ``irregular" has been used in the literature to describe a wide 
variety of morphologies including peculiar and interacting systems. 
Following Hunter \& Gallagher (1986) there is, however, a group of normal,
non-interacting, and intrinsically irregular
galaxies. These are usually divided into the Magellanic-type Ims, for which the
morphological prototypes are the Magellanic Clouds and into the smoother, 
IOs, which are usually referred to as amorphous 
IGs. Table~1 shows that the galaxies studied here belong predominantly to the 
Im type. The term ``dwarf galaxy" has been applied to galaxies fainter than 
M$_B$=-16 (e.g., van den Bergh 1966, Hodge 1971); and the data in Table~1 
suggest that in that sense, all of the galaxies studied here are dwarf 
irregular galaxies (dIs). We note that our sample galaxies are rather 
similar to
the Local Group dI IC~1613 (Freedman 1988) in terms of absolute 
blue magnitude and metallicity.

The observational details of 11 dIs studied by us with CCD photometry in
Johnson B and Cousins R were given in Hopp \& Schulte-Ladbeck (1995, HS). 
The observations consist of a uniform sample of 10 dIs (i.e., observed 
in the same run, calibrated with the same standards) and another run in 
which only the galaxy DDO 210 was observed. Since DDO~210 was not observed 
with the same set-up and the same calibrations as the other dIs, 
we exclude this galaxy from our comparison. We presented preliminary 
results on DDO~210 in Hopp \& Schulte-Ladbeck (1994). 
The seeing conditions for the uniform-sample 
observations ranged from about 1" to 2", and the limiting magnitudes 
reached at 
a DAOPHOT error of 0.1$^m$ were between 22$^m$ and 23$^m$, with the B 
images usually going about 0.5$^m$ to 1$^m$ deeper than the R images 
(see HS Tables 1 \& 2). We note that the total magnitudes derived by us 
(to the 26.5$^m$/arcsec$^2$ isophote, HS Table 3) are in excellent agreement 
with those given by KK (quoted to be in the B$_T$-system
of the Second Catalog of Bright Galaxies, RC2, which uses the 
25.0$^m$/arcsec$^2$ isophote). The difference between our total blue magnitudes 
and KK's is (-0.07$\pm$0.11)$^m$. We take this good agreement to indicate 
that there are no large systematic errors in our calibration. We also compared
single-star photometry of two galaxies in the uniform sample, UGC~8024 and 
UGC~8091, with previously available photometry by other groups of authors. 
The comparison revealed that the difference in the zero-point calibrations 
of single-star photometry is smaller than 0.2$^m$ in B, and that the colors
agree very well.

The goal of this paper is a comparison of the stellar content of the 10 dIs. 
Our results for UGC 5272 (which we found to have a small, neighboring dI, 
hence the nomenclature UGC 5272 A,B) were already published in 
Hopp \& Schulte-Ladbeck (1991). In the present paper then, we shall 
examine the properties of an additional 8 dIs from the same observing run. 
We wish to emphasize that whenever we are comparing the properties  
within the uniform sample, the results will be less affected by errors in
the calibration. Therefore, the relative properties of these galaxies 
will be more reliable than their absolute properties.

\section{H$_{\alpha}$ imaging}

In addition to the B and R images introduced in HS, we also obtained
H$_{\alpha}$ images of several sample galaxies between March 6 and 8, 1988. 
These data were taken during an instrument test run at the prime focus 
of the Calar Alto 3.5-m telescope, with the focal 
reducer, and a 371 by 561 pixel GEC chip which did not have the
best cosmetic properties. The images have  
a scale of 0.37~arcsec per pixel. We used an H$_{\alpha}$ filter 
centered at 6580~\AA\ with a width (FWHM) of 100~\AA. 
We employed an I-band image centered at 8700~\AA\ and with a 
width of 1200~\AA\ to accomplish the continuum subtraction. 
Exposure times and seeing values
(FWHM) are given in Table 2. The resulting H$_{\alpha}$-I images are 
displayed in section 4. 

After the usual CCD corrections, the I images were scaled to the
H$_{\alpha}$ frames by measuring the fluxes of several well 
isolated stars on
both frames, dividing the I frames by the flux ratio, and 
substracting them
from the H$_{\alpha}$ images. The electronics of the camera and of 
the chip produced unexpected, remnant over-exposure features 
and read-out bugs. They are easily visible in the images. Given the sharp 
edges of the CCD bugs, it is unlikely that they would be confused 
with real features in the galaxies. We shall therefore use 
the data to discuss the H$_{\alpha}$ morphologies of the sample
galaxies, but shall not give absolute fluxes. Similary, I frames
were not used to conduct single-star photometry; owing to the bad
seeing the galaxies are also not well resolved in I. No I frames could
be taken for UGC~6456 and UGC~8320, and the H$_{\alpha}$ images shown
therefore still contain a continuum contribution.

While H$_{\alpha}$ data exist in the literature for most of our galaxies 
(e.g. Hunter et al. 1993, HHG), we present the first published image for 
UGC~8024. This galaxy has attracted much interest owing to its dark-matter 
halo (Carignan \& Beaulieu 1989).
In addition, most of our images are deeper than those presented by HHG and add
some interesting new morphological details like the giant shell in UGC 7559.

\section{THE RESOLVED STARS}

The B \& R single-star photometry data were corrected for galactic 
foreground extinction, using the values given in Table 3 of HS. 
The galactic extinction law used was that of 
Cardelli, Clayton \& Mathis (1989), and the value for the 
total-to-selective extinction, R$_V$, used was 3.1. 

We begin our presentation of the resolved stellar content with a set of
three Figures for each galaxy. In each set, there is an illustration of the 
resolved objects identified with DAOPHOT in the B images. The 
brightness-coded position plots can be compared with the R images reproduced
in HS; they have the same orientation, S is up and W is to the left. 
Dashed lines on the B charts were drawn to separate the
area encompassing the galaxy from surrounding areas on the chip which
are thought to contain mostly galactic foreground stars as well as a few
distant and unresolved galaxies. The very brightest stars were ignored
by setting the saturation switch in DAOPHOT to 55,000 CCD counts. 
In some cases, DAOPHOT identified extended objects that look 
like background galaxies on the original images with single ``stars'' 
or chains of ``stars''; these are labeled on the B charts. The second set
of plots is a color-coded position plot. For each galaxy, the bluest objects
are plotted with the largest symbol, yellower objects are shown by using
an intermediate-sized symbol, and the reddest objects are displayed with
the smallest symbol. The B-position plots, color-coded position plots and 
the third set of plots, the color-magnitude diagrams (CMDs), use the same
plot limits and the symbols are coded in the same way, thus facilitating
a comparison of the data between different galaxies. 

The CMDs, given both for B vs. B-R and R vs. B-R,
show, plotted in separate panels, all stars found in both B and R, 
the foreground stars and the galaxian stars. Of course, this is not 
intended as a rigorous distinction of the foreground vs. galaxian stars, 
but merely as an illustration of the properties of the 
foreground contamination. Our procedure to select stars belonging to
the galaxies is a trade-off: CCD images of areas at larger distances from 
a galaxy are needed, but can in reality rarely be obtained due to the pressure
on observing time at large telescopes (cf. also Tolstoy 1995). Since we use
on-chip areas to learn about the foreground stars, and since the sample
galaxies will not have sharp boundaries to the distribution of
the resolved stars, we are very likely to include some stars that belong 
to the respective galaxy in the foreground contribution. On the other hand,
the CMDs labeled ``Stars in the Galaxy'' will sometimes show a few
very bright objects, clearly foreground stars which happen to be projected 
onto the area of the galaxy. Of course, these objects were ignored in 
interpreting the CMDs of the sample galaxies. The on-chip data 
indicate up to tens of objects in the foreground over the area of the
CCD chip.

Lacking galaxy-by-galaxy filed-star corrections, we applied the following
statistical approach to understanding the contaminations. We combined into one
CMD the data for the field stars of all sample galaxies. 
Using the combined field-star data, we
calculated how many field stars we should expect within the area of a given
sample galaxy (i.e., within the borders outlined on the position plots), 
and in which color and mangitude intervals they would
be most prominent. We find that 
corrections (of typically 5 stars) are predicted primarily in the 
color-magnitude range of 23$<$B$<$22 and 0.5$<$B-R$<$2.0; in all other
areas of the CMDs the corrections are very small.

An alternative way to obtain an estimate of the expected foreground 
contamination is to use a model of the number of Galactic stars (i.e., Bahcall
\& Soneira 1980). This method has its problems since the Bahcall
\& Soneira model does not always agree with observations (Kraft 1989).
The total number of Galactic stars predicted from the model in the B 
filter between magnitudes 16 and 24 towards the sample 
galaxies is between 10 and 20, in agreement with the data (but note the
small number of stars involved in both the data and the model).  

The interpretation of the CMDs depends very much on both the incompleteness of
the data, as well as on the distances of the galaxies studied. In measuring
stars at fainter and fainter magnitudes in a sample galaxy, we miss larger
and larger fractions due to incompleteness. To model
the incompleteness of the data, we added test stars covering the same range
of magnitudes as the original detections, to the B images
of those galaxies for which we detected more than 100 stars. We then
used DAOPHOT to recover the test stars. These simulations show that we
expect 100\% completeness to the 21$^s{}^t$ magnitude, and 90\% completeness
to the 22$^n{}^d$ magnitude. For all of the sample galaxies, the brightest
member-stars are found to be at B$_o$$\leq$21 (see Table 4). The simulations were 
carried out with test stars scattered uniformly across the sample galaxies. It is 
expected that the incompleteness in the detection of luminous blue stars in star 
clusters or HII regions is actually higher than simulated (``R136 effect").

The interpretation of the CMDs also depends on the effects that crowding
has on the magnitudes and colors of sources which are assumed to be single 
but are not.
In our discussion of the CMDs of individual sample galaxies an issue that
will come up repeately is how to interpret bright,
yellow objects. Although crowded objects can be de-blended in DAOPHOT using a 
pre-defined point-spread-function, one anticipates that this procedure
eventually has its limits in very crowded regions. This will
lead to intermediate colors as the light of red stars is mixed with that
of blue stars (Greggio et al. 1993). Gallart et al. (1996)  present an
in-depth discussion of crowding and find in particular that the distribution
of magnitude- and color-shifts at any given magnitude or color index is not
Gaussian. There are a few galaxies in our sample which show very outlying, 
blue data points on their observed CMDs. They can probably be explained 
with the large tails
seen at faint magnitudes and extreme color indices in crowding
simulations, or they could be background galaxies which we mis-classified
as stars.

Another explanation for the bright, yellow point sources is that they
are unresolved star clusters in the sample galaxies. Bica et al. (1990)
published colors for populous, young and intermediate-age clusters, whose
colors would occupy the areas of concern in our CMDs. Note that the Bica 
et al. results are based on empirical data, i.e., cluster spectra which 
hence include the contribution of H$_{\alpha}$-line emission to the R filter. 

In order to assign approximate spectral types from the B-R colors, we
used the Johnson (1966) Tables, after transforming the R magnitudes to
the Cousins system using Bessel (1983). These apply for Galactic
stars; the colors of stars in low-metallicity galaxies will be
shifted to earlier spectral types. The Johnson Tables do not contain data for early O-type
stars. When we measure colors that are more extreme than those
listed for the latest O-type stars in Johnson's Tables, we 
identify such an object as an O star. We note that the
colors of O stars are degenerate (e.g., Massey et al. 1995).

In the absence of primary distance indicators, distances to resolved 
galaxies have been estimated using the brightnesses of the three 
brightest blue and red supergiants as distance indicators 
(e.g. Hubble 1936, Sandage \& Tamann 1974, Humphreys 1983). A recent 
summary of avialable data was given by Karachentsev \& Tikhonov (1994).
The method has a number of problems (e.g., Humphreys \& Aaronson 1987). 
Since massive stars are usually found in associations, these could 
be mistaken for the brightest blue stars. Discriminating the brighest red 
supergiants against galactic dwarf stars of the same color is 
extremely difficult, and so they too are problematic distance indicators. 
Both blue and red supergiants are usually variable stars 
(e.g., Sandage \& Carlson 1985). 

An excellent summary of the problems 
that observers are faced with when attempting to apply the method of the 
brightest supergiants, as well as calibration 
relations that use apparent (i.e. directly observable) magnitudes, 
were recently published by Rozanski \& Rowan-Robinson (1994).
Rozanski \& Rowan-Robinson present calibrations which make use of the 
well-established observation that 
there is a dependence of the luminosity of the brightest stars
on the luminosity of their parent galaxy. Significant scatter about these 
relations (presumed linear) however results in a
distance-modulus error for any given galaxy of no less than 0.55$^m$.  

The dependence of the luminosity of the brightest stars on galaxy 
luminosity has been explained as a statistical effect. For the blue supergiants, 
numerical simulations with a constant star-formation rate qualitatively
re-produce the above relation as well as the significant scatter
observed about it (e.g., Schild \& Maeder 1983, Greggio 1986). 

In section 6.1, we use the calibrations by Rozanski \& Rowan-Robinson to
derive distances for the sample galaxies. To do so, we had to adopt a philosophy
for how to find the apparent blue magnitude of the brightest blue
supergiants, and we wish to outline our approach before discussing the
actual CMDs. As  we go to fainter and fainter 
magnitudes into the observed CMD of a galaxy, we
must eventually encounter its stellar content, and so separate it from
foreground contamination. Due to crowding and increasing
errors for fainter magnitudes, the main-sequence/blue-supergiant plume
and the red-supergiant region of the ``Stars in the Galaxy" CMDs 
exhibit wide scatter. While it may be difficult to identify correctly the three individual,
brightest blue or red supergiants (foreground contamination, crowding, blending in
color), it may nevertheless be possible to define an upper envelope to the stellar 
contents of a sample galaxy, presumably comprised of its supergiants. In a
B, B-R CMD such an upper envelope has a large negative color slope,
whereas in an R, B-R CMD, the envelope is nearly constant, especially for
the red supergiants.
 
To illustrate this supergiant envelope, we show some template CMDs in
Figure 1, constructed using the absolute visual magnitudes 
of stars in the MK system as listed by Schmidt-Kaler (1965) together 
with the colors given by Johnson (1966), and transforming the
R magnitudes into the Cousins system using Bessel (1983). 
The difference in the color of the envelope slopes in B, B-R vs R, B-R 
can readily be recognized; sliding the CMDs of Figure 1 onto the
``Stars in the Galaxy" CMDs of the sample galaxies helps guide the 
eye in determining the apparent magnitude at which the
transition from foreground stars into the supergiant stars of a 
sample galaxy occurs. This difference in color slope of the
envelopes can also be reproduced using stellar evolutionary models.
For instance, adopting the Bertelli et al. (1994) isochrones we
investigated the theoretical supergiant envelopes for their
Z=0.001 model (to offset from the Galactic metallicity, Z=0.020,
implicit in the empirical data of Schmidt-Kaler) and for three
different ages between 10 and 100~Myr. The feature of the different
color slopes in M$_B$, B-R vs. M$_R$, B-R is clearly seen at all ages; the older the
stars, the lower the envelopes in terms of absolute magnitude, as
expected. In section 6.1, we discuss further how the
brightness of the brightest supergiants is translated into a 
distance estimate.

The distribution of stars on the CMDs of the sample galaxies indicates that 
red giants were below the detection limit in all cases. The tip of the 
first-ascent red giant branch (TRGB) has been seen in the CMDs of
resolved galaxies, and a method using V,~I photometry of the TRGB 
to determine galaxy distances was developed by Lee et al. (1993).
Since we cannot identify the TRGB in our data, we cannot apply this
method as a distance indicator.

\subsection{UGC 4459.}

The B brightness-coded position plot is displayed as Fig. 2. Fig. 3 illustrates
the location of objects of different colors. Both Figures
show a higher concentration of stars, and a preference for blue and yellow 
stars, in the area marked as belonging to the galaxy.

Inspection of the CMDs (Fig. 4) and the frequency plots reveals that
objects brighter than 20$^m$ in either B or R are likely to be foreground
stars. In particular, the brightest object on the chart, which is
also fairly red (object number 17, see HS), is concluded to be a foreground star 
fortuitously projected onto the face of the galaxy. The brightest blue
stars in UGC~4459 start to appear at B$_o$$\simeq$20.5$^m$ and 
B-R$_o$$\simeq$-0.2$^m$. Notice that there is no contribution to this portion of
the CMD expected from galactic foreground stars (Fig. 4). The colors of the
blue objects correspond to a population of early B supergiants in UGC~4459,
capable of ionizing their surroundings. Since the effect 
of H$_{\alpha}$ emission on the R filter leads to redder observed colors,
these blue objects could actually be of even earlier 
spectral types than those which are indicated by the observed
colors.  A V vs. B-V CMD of UGC~4459 was published by Karachentsev 
et al. (1994), who also suggest that the brightest blue supergiants start to
appear at around 20.5$^m$ in B.

The spatial distribution of stars in the R image (HS) as well as in the 
B chart (Fig. 2) shows two elongated, parallel features
which appear more densely populated than other areas across the face of 
the galaxy. The gap is unlikely to be due to absorption, i.e., a dust lane.
Both the R and the B image indicate that fewer bright stars are found
in the gap than in the features, while the color map (Fig. 3)
shows that blue stars are being detected in the gap. 
We therefore interpret the two features as being due to
higher stellar densities in these regions of UGC~4459. 

Our H$_{\alpha}$ image is shown in Fig. 5. HHG also published 
an H$_{\alpha}$ image of UGC~4459; the two agree very well. Another
H$_{\alpha}$ image is avaiable from Strobel et al. (1990). 
The overall morphology of their image also compares well with
our's. Fig. 5 shows several bright HII regions which coincide spatially 
with the two areas of high stellar density. There is also faint 
and diffuse emission in this galaxy as well. The bluest stars with 
(B-R$_o$$\le$-0.1$^m$) are all located in the areas outlined by 
the H$_{\alpha}$ emission. We count 14 HII regions in UGC~4459 (see Table 4). 
Strobel et al. distinguish 18 HII regions in their image. Of course, in both
cases, there is an arbitrariness in the object identification and
where the boundaries of HII regions are drawn. For instance,
our image does not suggest the large number of HII regions
Strobel et al. identify as regions 1 to 7, and instead displays more
detail in regions 10 and 11.  

The B vs. B-R CMD shows a number of bright, yellowish objects, with 
B$_o$$\simeq$21 and colors in the approximate range of 
0.2$^m$$\le$B-R$_o$$\le$0.5$^m$.  There seem to be no foreground objects 
having similar properties. The yellow objects are located in the same regions 
across the face of UGC~4459 as the blue stars. Three possible explanations
present themselves. First, the bright and yellow objects could be an 
artifact of the reduction. Since they are situated in the 
densely populated regions of the galaxy, crowding is a problem. 
Second, populous clusters of young to intermediate age ($\le$2x10$^8$yr) 
produce colors in this range (0.3$^m$$\le$B-R$_o$$\le$0.4$^m$, 
Bica et al. 1990). The brightest
of the yellow objects could therefore be interpreted as star clusters. 
In the case of UGC~4459, we do not believe this to be the case often, since
the brightest yellow objects seem to be part of a distribution of stars
on the CMDs that extends smoothly to fainter magnitudes, and they also
fit in with the overall brightness level of potential blue and red 
supergiants. Third, the yellow supergiants could be the evolutionary 
descendants of the most recently formed massive stars in UGC~4459.

The red objects with B-R$_o$$\simeq$2$^m$ and R$_o$$\le$20$^m$, 
although severely ($\sim$50\%) contaminated by galactic foreground 
stars, may contain a 
contribution by red supergiants in UGC 4459. In spite of the fact 
that we are dealing with a very small number of objects, 
this can be seen by comparing the B vs. B-R with the R vs. B-R CMD. 
The upper envelope to the most 
luminous stars in the B vs. B-R decreases by about 2.5$^m$
from blue to red, whereas that in the R vs. B-R remains quite level. 
This is the expected behavior for Ia supergiants (compare Fig. 1). 
Using the luminosity and color calibrations cited previously, we find 
that red giants in UGC 4459 are below the detection limit.  
Furthermore, the colors of the bluest stars and the detection limit 
rule out that our data reach the main-sequence for this galaxy.

\subsection{UGC 7559.} 

Figures 6 to 9 display the data for UGC~7559.
 
The two brightest objects (objects 133 and 112, see HS) in the B vs. 
B-R CMD of stars in the galaxy (Fig. 8) coincide with regions of 
HII emission (Fig. 9). This might explain their yellowish colors with   
H$_{\alpha}$ contribution to the R filter. An H$_{\alpha}$ 
image of UGC~7559 has also been provided by HHG, but our image shows 
more detail. For instance, not detected by HHG but clearly visible 
on our image is a super-shell 
measuring $\sim$27". Several blue stars are coincident with 
the shell region (Fig. 7). We count a total of 12 HII regions on our image. 

The R vs. B-R CMD of the galaxy shows the tip of the distributions
of objects which might be interpreted as the brightest, single
yellow and red supergiants in UGC~7559 at R$_o$$\simeq$19.5$^m$-20$^m$. 
This is consistent with the brightest blue stars being found at 
B$_o$ and R$_o$ of just above 20$^m$. There is a gap in the B vs. B-R
CMD in the brightness distribution of the brightest blue supergiants 
between the
brightest object at B$_o$$\simeq$20$^m$ and the remainder of the blue 
(B-R$_o$$\le$0$^m$) population, which continues below B$_o$$\simeq$21$^m$. 
 
The spatial location of the stars with colors around B-R$_o$=0$^m$ is
along the perimeter of the face of the galaxy as seen on the R image.
However, the main ``blue" supergiant plume in this galaxy is quite yellow,
with B-R$_o$$\simeq$0.5$^m$. These yellow supergiants with colors 
between 0$^m$ and 
1$^m$ are distributed much more uniformly across the face of the galaxy than 
the bluer objects. They also populate an outlying
association which can be seen on the B chart at coordinates of X=10" and
Y=90", and which on the R image of HS looks disconnected from the light 
distribution of the main body of UGC~7559. As Figures 7 and 9 show, 
this association coincides with an area of H$_{\alpha}$ emission. 
Outlying associations or HII regions are also 
noticable in other galaxies in this sample (see below). 

Since the galaxy subtends a large area of the CCD chip, what is displayed as 
foreground contamination in the CMDs is even more uncertain for this than 
other sample galaxies. Bearing this in mind, we infer from the CMDs that 
there is a substantial red supergiant population in UGC~7559. 

\subsection{UGC 8024.}

Figs. 10 - 13 feature our data for UGC~8024.

Optical images as well as an HI map of UGC~8024 were published by 
Carignan \& Beaulieu (1989). The inner
HI-surface-density isophotes display a triangulary shaped morphology that is
similar to the shape of the body of the galaxy in the optical images. The HI
envelope of this galaxy is found be to very extended, about 5 times the Holmberg
diameter. In order to explain the rotation curve, a dark-matter halo is needed.

The galaxy has been well resolved into single stars by our observations, 
and the CMDs (Fig. 12) show a
high contrast in the numbers of stars within the galaxy to those observed
outside of the boundaries drawn in Fig. 10. The envelope of the brightest 
supergiants is well defined, and there is little concern about confusion with 
galactic foreground stars. The supergiants of UGC~8024 appear at magnitudes of 
around 20$^m$. Carignan \& Beaulieu (1989) also obtained single-star
photometry of UGC~8024; they resolved 25 stars. They find 
$<$B(3)$>$=20.34$\pm$0.24 for the three brightest blue supergiants.
The blue plume of the CMD in Fig. 12 is very well populated. 
The mean color of the plume is $\sim$0.1$^m$, typical of A-type stars. 
There is also a fair number of stars with colors of G-to-K supergiants.

A few data points, which have large error bars, are found at 
B-R$_o$$\simeq$-0.7$^m$. Several of these are well above the detection limit, 
and so we do not believe their colors to be artifacts of the measurement
process. The colors either suggest several mis-identified, very
blue background galaxies, or they can be interpreted as being 
indicative of the presence of O-type stars in the galaxy. Since 
the magnitudes of these objects 
start at 2.5$^m$ below that of the envelope of the brightest supergiants, 
they would include main-sequence O stars. The stars in the blue plume are 
located all across the face of the galaxy. Even bluer
objects with B-R$_o$$\le$-0.2$^m$ also follow this distribution. 

Our H$_{\alpha}$ image of UGC~8024 is displayed in Fig. 13. 
The Figure shows about eight distinct HII regions, and diffuse emission from 
across the entire face of the galaxy. A prominent HII region is found to be 
located outside of the main body of the galaxy as defined by the R (HS) and
I (Fig. 13) images. The R and I images mainly show a 
triangularly shaped body of the galaxy. Faint resolved stars
are detected curving from the tip of the body to the northeast towards 
this outlying HII region. Diffuse H$_{\alpha}$ emission is seen in the 
H$_{\alpha}$ image to extend to the northeast of the tip of the main body as 
well. This area of emission coincides with several
blue objects near X=110", Y=135" in Fig. 11, suggesting that it 
might be another case of an outlying association.

\subsection{UGC 8091.} 

Our data for this galaxy are diplayed in Figs. 14 through 16.

According to Hodge et al. (1989) the center of this galaxy is
dominated by HII regions; they count 32 individual HII regions. 
The H$_{\alpha}$ image of HHG shows only one, 
maybe two, fairly weak HII regions. HHG classified one as a possible shell. 
This shell is clearly visible on the deep R frame of HS. 
Drissen et al. (1993) searched UGC~8091 for the presence of Wolf-Rayet (WR)
stars, the descendants of the most massive O stars, and found none.
They suggest that this is due to the low massive-star content of this galaxy.
In other words, the lack of WR stars could be a statistical effect.
We further note that at low metallicity, only a small fraction of massive 
stars is expected to enter the WR phase; and the WR phase 
is also predicted to be a short stage in the evolution of massive stars  
-  at ages of 7-10 Myrs the WR stars of a coeval population 
disappear while there are still low-mass O stars that can produce nebular 
emission (Meynet 1995).
    
There is a plume of blue stars with colors B-R$_o$$\simeq$-0.1$^m$, 
equivalent to mid-B supergiants. The frequency plots and the 
CMDs (Fig.~16) indicate that most of the
brightest objects are members of the galaxy. If the brightest supergiants 
are those objects with R$_o$ and B$_o$ of about 19$^m$, we may expect 
to see some early-type main-sequence stars in the CMDs. According to Fig.~16, 
for these objects, confusion with faint blue foreground stars
is also likely. The plume of blue supergiants is much narrower 
in color than in some of the other sample objects,
probably due to the fact that this object is closer and better resolved,
hence the effect of yellower colors resulting from insufficient
de-blending of the stellar images is not so severe. The frames of 
this galaxy are also those with the best seeing (see HS, Table 1) 
in the sample. The bright 
objects with colors B-R$_o$$\simeq$2$^m$ are considered to be
the red supergiants of UGC~8091. 

A number of previous studies of the stellar content of UGC~8091 based on
CMDs exist, however, all of them use different color systems than 
HS (e.g., Hoessel \& Danielson 1983, de Vaucouleurs \& Moss 1983, 
Aparicio et al. 1988, Tolstoy 1994, Tolstoy 1995). The data presented 
here do not go to deeper limiting magnitudes than the ones published 
before, so they merely provide a consistency check. Tolstoy
et al. (1995) derive an extinction corrected Cepheid distance 
modulus for UGC~8091 of (26.75$\pm$0.35)$^m$, based on the 
identification of one Cepheid variable. We see the brightest blue 
supergiants at a B magnitude of 19.0$^m$; Hoessel \& Danielson
find the brightest blue supergiants at B$\leq$19.5, and Aparicio et
al. give $<$B(3)$>$=18.88. We derive a distance modulus based on 
the blue supergiants in section 6.1, and it is in agreement with 
Tolstoy et al.'s value.

Tolstoy (1995) discusses the morphology of neutral gas (HI), ionized gas (HII)
and of blue and red stars in UGC~8091. She finds that red stars are spread
uniformly across her UGC~8091 images, whereas blue stars and HII regions
lie more in the center of the galaxy and follow the peaks in the HI distribution.
From the interpretation of the CMDs she suggests a declining star-formation rate
over the past 1 Gyr.

\subsection{UGC 5272 A and B.}
   
The B-filter frequency plot and the chart (Fig. 17) show that 
there are several bright blue objects that are 
located within the body of UGC 5272~A and which have no counterparts of similar
brightness in the foreground. In the B vs. B-R CMD (Fig. 19), these show up as  
four bright objects with colors around 0.4$^m$. 

We have argued (Hopp \& Schulte-Ladbeck 1991) that at least three of 
these objects (objects number 43, 23, and 80, see HS) could be star 
clusters in UGC 5272A. They coincide with the location of HII regions, which
suggests that they may be the ionizing clusters. (Note that HHG presented 
an H$_{\alpha}$ image of UGC~5272 showing faint emission from the position 
of UGC~5272~B as well, which was located outside of the H$_{\alpha}$ frame
of Hopp \& Schulte-Ladbeck 1991). There are few resolved stars in that galaxy.
Hopp \& Schulte-Ladbeck (1991) assumed that the fourth object, which is also 
the fourth brightest blue
object in UGC5272~A, could be a bright blue supergiant and used it in
estimating a distance from the method of the three brightest blue supergiants. 
After also comparing the blue supergiants in this galaxy with those of the LMC,
they decided to adopt a distance of about 6 Mpc, much closer than that indicated
in KK. (The estimate derived in section 6.1 puts UGC~5272~A at an even 
smaller distance.)

The frequency plots suggest with some certainty that the galaxy population in 
UGC~5272~A sets in below 20$^m$. The uncertainties both in terms of error bars 
and foreground confusion are large in the red part of the CMDs (Fig. 19). This makes it 
difficult to assess the number of red supergiants. Our data indicate that the 
number of red supergiants in UGC~5272~A is small. The bluest stars, 
B-R$_o$$\le$0$^m$, are faint, starting at magnitudes of around 21.5$^m$. Several of these 
stars cluster in the outlying association labeled in Fig. 1 of Hopp \& Schulte-Ladbeck 
(1991) and located near X=150", Y=130" on the charts (Figs. 17 \& 18).
The B vs. B-R CMD shows a prominent, bright plume 
of yellow supergiants, which reaches up to about B$_o$$\le$20.7$^m$ or 
R$_o$$\le$20.2$^m$. These objects are likely to contain additional blue 
supergiants, which appear at redder colors due to the effects discussed 
above (blending, H$_{\alpha}$).

UGC 5272 is considered a blue compact dwarf galaxy (BCD, also known as isolated 
extragalactic HII regions) in the list of Kunth \& S\`{e}vre (1986). This is also
the case for several other of the sample galaxies, illustrating perhaps the
absence of a firm definition that distinguishes BCDs from dIs.

\subsection{UGC 5340.}

Our data for UGC~5340 are presented in Figs. 20 - 23.

The R image of HS shows that this galaxy is poorly resolved. 
The CMDs (Fig. 22) of the foreground stars indicate that the 
contamination is severe for the brightest objects in UGC~5340. There are 
few stars in the CMDs which are actually expected to be objects in UGC~5340.
It is therefore difficult to decide at which magnitude level the brightest 
supergiants occur. According to the frequency plots, stars in the galaxy 
are certainly found at magnitudes larger than 21$^m$. Most of them are rather
blue. It is difficult to detect a genuine red supergiant population. 

An H$_{\alpha}$ image of UGC~5340 was published by HHG. Our image is
shown in Fig. 23 and is qualitatively in agreement with HHG, but at higher
signal-to-noise it shows more detail. In addition to numerous distinct HII 
regions also seen on HHG's image, Fig. 23 suggests that there is diffuse
emission in the central regions of UGC~5340 as well. The CMDs indicate
a small population of blue objects in the center. A comparison
of the star chart in Fig. 20 with the R image of HS shows that many of the
detected stars are situated in an outlying patchy extension at the northern end
of the main body of UGC~5340. This extension contains some of the bluest stars
and - according to the H$_{\alpha}$ images - at least 3 bright and 3 or 4 faint 
HII regions. At the southern end of the main body, another extension is visible 
in the R frame of HS which is also associated with faint H$_{\alpha}$ emission, 
but we detected no stars inside this region. 

UGC 5340 is listed as a type Im/BCD in the catalog of Binggeli et al. (1990). 

\subsection{UGC 6456.} 

Figs. 24 to 27 show our data for this galaxy.

Very few objects are detected in UGC~6456. The R image of HS reveals that
this galaxy is not well resolved. While the integration times
were as long as those for most other sample galaxies, this was 
an observation taken under poor seeing conditions. 
The frequency plot indicates that only blue objects 
are seen within the area considered to belong to the galaxy. 
A comparision of our B-Chart with our H$_{\alpha}$ image suggests that
these blue sources coincide with regions of strong H$_{\alpha}$ emission. 
The blue objects are found in a small region of about 20" in diameter which 
is offset from the center of the galaxy in the R-band image of HS. 
Thuan et al. (1987) counted 8 HII regions in UGC~6456.
Fig. 27 indicates 5 bright, and perhaps another 5 faint HII regions.
That this galaxy is very actively forming stars is 
indicated by the recent detection in X-rays of a hot gas outflow
driven by the present starburst (Papaderos et al. 1994)

The morphology of UGC~6456 resembles that of UGC~8091. UGC~6456 has 
also been classified as a BCD (Kunth \& S\`{e}vre 1986). It is considered 
the nearest BCD, and amongst the BCDs, belongs to the morphologically most 
frequent type of ``iE" galaxies, 
which show elliptical outer isophotes and several areas of active star 
formation distributed irregulary near, but offset from, the center  
(Loose \& Thuan 1985). 

A V vs. B-V CMD was previously published by Karachentsev et al. (1994), 
but we are unable to
match their chart of resolved stars up with our's. This could be due 
either to the poor quality of their chart or the small number of
point-sources identified in our frames. Since several of the bright objects
we identified as point-sources and plotted on the CMDs coincide spatially 
with areas of very strong H$_{\alpha}$ emission, they may me unresolved 
stellar associations rather than single stars. This, and the small
number of sources has prevented us from deriving a supergiant-envelope
based distance for UGC~6456. We are currently in the
process of analyzing HST images of this galaxy, which
are of much better resolution than the available ground-based images and
allow us to produce better CMDs for this galaxy (Schulte-Ladbeck et al. 1998).

\subsection{UGC 8320.}

Our data for UGC~8320 are displayed in Figs. 28 through 31. 

The two brightest objects in the CMDs (Fig. 30) are considered to be foreground 
stars which happen to be projected onto the face of the galaxy. The next two  
brightest blue objects are identified as members of UGC~8320, 
owing to their association with HII regions. 

An H$_{\alpha}$ image of UGC~8320 was published by HHG, and our image is shown
as Fig. 31. On either image, UGC~8320 displays a wealth of bright HII regions, 
which are found distributed across the entire face of the galaxy. This suggests
that the entire galaxy is presently in a state of vigorous star formation. 
There is a chain of regions forming a ridge across the center to the north-west 
edge of the body of the galaxy as seen in the R image of HS. Most of the 
resolved blue objects are distributed along that ridge as well (see Fig. 29). 
A few bright HII region are found near the southern tip of the galaxy and 
also include a blue object in our single-star photometry. A group of stars 
near X=80", Y=80" on Fig. 29 does not match up with any HII regions.

We assume that the blue supergiants have B$_o$$\simeq$19.5$^m$. 
Whereas there is a plume of blue supergiants, very few stars that could be red 
supergiants in UGC~8320 are seen. The bluest stars in UGC~8320 have colors 
which are consistent with those of O or early B. The CMDs may reach the 
main sequence for these stars.

\subsection{UGC 8760.}

In Figs. 32 to 34, we present our data for UGC~8760.

We interpret the data to indicate that the supergiant component of UGC~8760 
appears at B$_o$, R$_o$$\simeq$20.5$^m$. There is a pronounced plume of blue 
objects in the B vs. B-R CMD (Fig. 34), with B-R$_o$$\simeq$0$^m$ and no 
serious foreground confusion. For objects with reasonable error bars, the blue
plume extends to B-R$_o$$\simeq$-0.5$^m$. Three of six objects with bluer colors  
are also at the detection limit of our photometry. Thus, these objects could 
either be interpreted as O-type stars, or they could be spurious detections 
(see above). The fact that all six very blue objects coincide spatially with
two areas of HII emission (HHG) at either end of the elongated body
of this galaxy argues in favor of them being early-type stars in UGC~8760. 
Considering the galactic foreground contribution, the CMDs can only have 
a small population of red supergiants belonging to UGC~8760.

Morphologically, the appearance of this galaxy on the R image of HS is that
of a very flat elliptical main body evenly populated with red stars. The blue
stars (see Figs. 32 \& 33) follow this distribution as well, thereby giving this galaxy
a not-so-irregular appearance.

\section{THE UNRESOLVED STARS}

An underlying light distribution is evident in the B and R images of 
all sample galaxies. In HS, we described in detail how we extracted these
smooth light distributions, in order to apply the PSF-fitting 
photometry discussed in the previous section. Here, we study the smooth light 
distributions which result from unresolved sub-threshold stars and - to
a smaller extent and mainly in the R filter - from the emission of 
gaseous nebulae (see the discussion in Hopp \& Schulte-Ladbeck 1987). 

While the inner parts of the light distributions of the unresolved stars
are often patchy, the overall appearance of the sample galaxies in both 
colors is relatively regular. They have elliptical shapes with only minor
variations in the amount of ellipticity and orientation as a function
of radius. We were able to follow most of these distributions out 
to a surface brightness, SB, of 
$\sim$27.0$^m$~arcsec$^{-2}$ in B as well as in R. Thus, the light 
distributions of the unresolved stars can be traced to twice the 
distances of the distributions of the resolved stars. 
The angular sizes, 2a$_{max}$~["], 
at which the brightness levels start to become indistinguishable from 
the background, are given in Table 3.

After smoothing the data with a spatial filter of a scale length 5",
we studied the regular part of the underlying light distribution. We 
applied the ellipticity fit of Bender \& M$\ddot{o}$llenhoff (1987) to the data.
Figs. 2 \& 3 of Hopp \& Schulte-Ladbeck (1991) show an example of 
the resulting surface brightness profiles. These fits demonstrate quantitatively 
that the amount of ellipticity as well as the orientation are either 
constant as a function of radius or vary only slightly. 
Mean values for $\epsilon$ are given in Table 3. The observed 
surface brightness profiles also yield values for the central
surface brightness in both colors, SB(0,fit) and the isophotal 
Holmberg diameter, 2a$_{H}$, in ["] (Table 3). A comparison with
Table 1 shows that the new sizes do not always agree with those
listed by KK.

While the ellipticity and the position angle of a galaxy are very similar in 
the two colors, the surface brightness profiles SB(a) are different. 
In particular, the color profile, SB$_B$(a)-SB$_R$(a)~=~$\Delta$SB(a), varies
with radius. In most cases, the profiles can be traced to larger distances
in R than in B and the galaxies are becoming redder toward larger radii. 

Most SB profiles are relatively regular, showing only small deviations 
from a linear behavior in SB over most of the spatial extension 
of the galaxies. (In some cases, central flattening and/or outlying 
cutt-offs exist). This kind of light distribution is well described 
by the exponential disk law. Within the errors, the values given in Table~3
of the scale length, $\alpha^{-1}$, are identical in B and R for 
most sample galaxies. In UGC~4459, 
UGC~8091 and UGC~8760, the value in R is slightly larger than that in B. 

\section{DISCUSSION}

Table 4 is a summary of derived characteristics of the sample galaxies.
The column headings are quite self-explanatory. The entries in columns
2-5 concern the supergiant luminosities and distances of the
galaxies, as described in detail section 6.1.

In column 6, we list the number of HII regions seen in each galaxy. This is
a subjective number since it depends on the depth of available exposures and the
stretch of the published images and should be taken as a lower limit to the
actual number of HII regions in a galaxy. We compared our numbers with those
given by other authors in the sections on individual objects.

We also give numbers and percentages for OB, AF, and GKM stars, which can be compared
with results of other authors, e.g., Hunter \& Gallagher (1985). 
They were derived by counting the numbers of stars
in the B-R color interval smaller than 0.0$^m$ for the OB stars, 
0.0$^m$ to 0.99$^m$ for the AF stars, and 1.0$^m$ and above for the GKM 
stars. Star counts in these color bins were carried out for both, stars in the galaxy 
and foreground star, samples. A correction was applied to account for the different 
areas on chip subtended by the galaxy and the foreground. UGC~6456 and UGC~5272~B 
have no entries, due to the paucity of data on
their resolved stellar content. For all of the entries on the numbers of
resolved stars, we are of course dealing with small number statistics.

In column 10, we list the B-R color of the underlying light distribution of 
unresolved stars as derived from the data in section 5. Finally, column 11 
provides an indication of whether or not an outlying association or HII region
is observed with a clear separation from the main distribution of stars in the 
respective galaxy.

\subsection{Supergiant Luminosities, Distances}

In Table 4, we give an estimate of the apparent blue magnitude corrected
for galactic extinction, of the envelope of the brightest stars which
we conclude to be members of the respective galaxy, B$^0$$_{env}$. 
As this step is quite subjective, both authors independently derived
a value for B$^0$$_{env}$ by inspecting the CMDs as well as the luminosity functions,
and the results agreed reasonably well.
Our method is accurate to $\pm$ 0.5$^m$ at best,
and so we give magnitudes rounded to the nearest half magnitude.
In Table 4, the objects are sorted by increasing B$^0$$_{env}$ and 
presumably, distance.

In order to determine distance moduli from the observed envelope
brightnesses, we used the calibration relations by Rozanski 
\& Rowan-Robinson (1994). Their Figure 10(f) relation gives a 
calibration for the apparent blue magnitude of the three brightest 
blue supergiants minus the total apparent blue magnitude of the galaxy, 
versus the total absolute blue magnitude of the galaxy. 
The total, extinction corrected, apparent blue magnitudes of the galaxies
were derived from the integrated B magnitudes given in Table 1 of HS, 
where the error is estimated to be about 0.1$^m$ 
(except for UGC~4459 where part of the galaxy was not included 
on the CCD chip). Note that the magnitude difference for UGC~8091 places it well below
the calibrated data range; we extended it with the regression.
The resulting M$^0$$_{B,T}$ values are listed in Table 4.
The photometric error for the apparent magnitude difference was 
derived by quadratically adding the supergiant-envelope error
(0.5$^m$) and B-magnitude error (0.1$^m$), and is
of order 0.51$^m$. Together with the error that Rozanski 
\& Rowan-Robinson list as that on a single observation
using the regression, 0.88$^m$, this leads to an uncertainty 
of 1.02$^m$ in the M$^0$$_{B,T}$ values listed in Table 4.

We then used Rozanski \& Rowan-Robinson's Figure 10(c) B-band
luminosity calibration to
derive the absolute blue magnitude of the brightest blue supergiants, 
M$^0$$_{B,BSG}$, see Table 4. The distance moduli can now be derived.
For the error in this calibration we 
adopted their minimum error on the distance modulus using the
regression, 0.90$^m$. The propagated minimum error on 
M$^0$$_{B,BSG}$ and the distance modulus is 1.36$^m$.

In our sample, a Cepheid distance exists only for UGC~8091 (Tolstoy
et al. 1995), m-M=(26.75$\pm$0.35)$^m$, yielding -12.43$^m$ for the 
extinction-corrected, total absolute blue magnitude of the galaxy.
Recall that our observed parameters for UGC~8091 lie beyond the extent of
Rozanski \& Rowan-Robinson's calibration relations but we assumed
the linear regressions continue to be valid for galaxies with 
lower absolute blue magnitudes. This predicts -6.93$^m$ for the 
B luminosities of the brightest blue supergiants, and 19.82$^m$ for
their apparent B magnitudes. We find the brightest blue supergiants at 
(19.0$\pm$0.5)$^m$. This is, within the errors, in agreement with the above
prediction, and in excellent agreement with the work others (Hoessel \& Danielson 1983,
Aparicio et al. 1988). We derive M$^0$$_{B,T}$=(-12.2$\pm$1)$^m$, which agrees
well with Tolstoy et al. The Rozanski \& Rowan-Robinson calibrations
yield M$^0$$_{B,BSG}$=-6.9 and m-M=25.9, with a minimum error of 1.36$^m$, 
values that agree with Tolstoy et al. within the errors.
Kharachentsev \& Tikhonov (1984) list M$^0$$_{B,BSG}$=-6.26$^m$, 
M$^0$$_{B,T}$=-10.4$^m$ and a distance modulus of 25.10 for UGC~8091. 
A small distance modulus seems unlikely because then the TRGB 
should be visible in Tolstoy's (1995) CMDs.
  
Our sample galaxy UGC~4459 is considered a member of the M~81 (or B2) group. The
Cepheid distance of M~81 is m-M=(27.80$\pm$0.20)$^m$ (Freedman et al. 1994).
We find the brightest blue supergiants of UGC~4459 at (20.5$\pm$0.5)$^m$, and
the resulting distance modulus is 27.8$^m$, in excellent agreement with
the M~81 distance. (Recall that the distance-modulus error is expected to be 
larger than the formal 1.36$^m$ error due to the unknown measurement error 
on the total apparent blue magnitude of the galaxy.) Kharachentsev 
\& Tikhonov (1984) give M$^0$$_{B,BSG}$=-7.36$^m$, M$^0$$_{B,T}$=-13.2$^m$
and a distance modulus of 27.66$^m$. 

For UGC~6456, Schulte-Ladbeck et al. (1998) used an HST/PC2 I, V-I CMD
to find a TRGB distance modulus of (28.4$\pm$0.09$\pm$0.18)$^m$. We can use
this distance estimate to evaluate at what apparent B magnitude we should
be finding the brightest blue supergiants in the calibration of 
Rozanski \& Rowan-Robinson. Using HS for the apparent blue magnitude
of the galaxy, 14.19$^m$, and the above distance modulus,
we find the absolute blue magnitude of UGC~6456 to be
-14.21$^m$. The Rozanski \& Rowan-Robinson Figure 10(c) 
B-band luminosity calibration then yields M$^0$$_{B,BSG}$=-7.43, 
indicating we should see
the brightest blue supergiants at an apparent B magnitude of
about 21.0$^m$. The HST data do not
include the B and R bands. The three brightest blue supergiants are found
at a V magnitude of about 20.1$^m$ with colors that correspond to
early A-type supergiants. This predicts that in the B-band, they
should become visible around 20.2$^m$ (recall the minimum error
from the Figure 10(c) regression is 0.9$^m$, so ground-based and HST results are not
inconsistent). In order to make another, more direct
comparison with the HST data, we may also use the  
brightest red supergiants, which appear at a V magnitude of about
21.2$^m$ in the HST CMD. We can now use Rozanski \& Rowan-Robinson's Figure 10(a)
relation to derive the absolute V magnitude of the brightest red supergiants
to be -7.08$^m$, and their apparent V magnitude should hence be 21.3$^m$.
Here, the agreement between the data and the Rozanski \& Rowan-Robinson
prediction is excellent.

A comparison of our distances with those of KK shows that we place 
UGC~5272~A and UGC~5340 about a factor
of two closer, from the 8 to 9~Mpc range to the 4 to 5~Mpc range.
Our results make sense since we resolve many stars in these galaxies.
We also find that UGC~8024 and UGC~8320 may be closer than
their KK distances. For UGC~6456, on the other hand, the HST TRGB distance is
twice that of its KK distance.

\subsection{Stellar Content and Star-Formation Histories}

As a note of caution before we start our deliberations on the stellar content 
of the sample galaxies, we point out that colors and magnitudes of
stars of different stellar masses and ages 
depend on metallicity, and that the interpretation of the data is sensitive to
the stellar models produced by different groups. We mean the age bins 
into which we grouped the discussion sections below as a rough division of ages
as indicated by the presence of certain stellar types and by the color of
the integrated light.

\subsubsection{Stars younger than 10Myr}

There are several indicators for the presence of very young stellar components 
in galaxies, namely direct observation on the CMD of main-sequence stars with 
masses of around 15~M$\sun$ and above, detection of H$_{\alpha}$ emission which 
also evidences massive, H-ionizing stars, observation of WR stars which 
represent a very short-lived phase ($<$3Myr) in the evolution of massive stars. 

The detection of WR stars is usually accomplished with either imaging or spectroscopy data
by discriminating the ``WR feature" at around 4650{\AA} against the continuum. 
The sample galaxies as a whole have not been surveyed specifically for 
presence of WR stars. The exception is UGC~8091 which was searched for WR stars 
with imaging, but none were found (see section 4.4).
Kinman \& Davidson (1981) combined spectra of several galaxies exhibiting similar spectra,
including UGC~5272~A, and detected a weak ``WR feature" in the resulting spectrum of
higher signal-to-noise. It is not clear, however, how much of this feature is
attributable to WR stars in UGC~5272~A. Absence of the ``WR feature" in UGC~6456 
indicates that no WR stars are present in this galaxy, at least at those positions
sampled by the spectrograph's entrance aperture (Izotov 1998). 
We did not find information on other of the sample galaxies in the literature.
Hence, our knowledge of star births in the last few Myr is limited. The observation of 
both blue and red supergiants in the CMDs suggests, however, that star-formation in the 
sample galaxies did not experience a recent, sharp temporal cutoff.

All of the sample galaxies display HII emission, clearly organized
into HII regions and coincident with young stellar associations. For several galaxies, we
thought the main-sequence was detected in the resolved stars as well. 
Following Maeder \& Meynet (1989), the H-burning lifetimes of OB stars 
are of the order of several 10$^6$ to 10$^7$ years.  
Hence all of the sample galaxies have made massive stars within the last about 10Myr.
The observation of both blue and red supergiants in the CMDs of the
sample galaxies suggests that star-formation was
active over a recent period of several 10Myr.

One of the sample galaxies deserves to be noticed. UGC~8320 clearly stands out
in a comparison of H$_{\alpha}$ images of the sample galaxies. It also
has the highest fraction of OB stars in the sample (although there does
not appear to be a correlation between the number of HII regions and the
fraction of OB stars among the sample galaxies in general). Its
background-light color is so blue that we suggest this galaxy has made
a significant fraction of its stellar mass in recent history.
 
The locations of the HII regions and of the young stellar associations show diverse 
morphologies among our sample of galaxies, ranging from being distributed all across 
the entire face of a galaxy, to being concentrated only in specific regions, sometimes 
near the center of a galaxy. This suggests that at a given time a variety of locations 
within the potential well of a dI can act as star-forming sites.

\subsubsection{Stars younger than 100Myr}

The presence of supergiants can be used to track star births into the
several 100Myr interval of the star-forming history of the sample galaxies. The
distances are crucial here to reveal how deeply we view into the stellar content
to fainter magnitudes and hence, lower masses and larger ages. Overlaying stellar-evolutionary
tracks onto our CMDs (e.g., Bertelli et al. 1994) we find that we can in general detect 
supergiants with ages of up to around 50Myr. Beyond 50Myr, the data are severely affected by 
incompleteness and blending.

With the usual caveats, we find that blue and red supergiants are detected
in all of the sample galaxies. (The ``non-detection" of red supergiants in UGC~8320 is 
most likely due to small-number statistics combined with an over-correction of the 
Galactic foreground. In other words, we do not claim that the data prove an absence 
of red supergiants in this galaxy.) In particular, we
do not observe the presence of red supergiants without the simultaneous presence
of blue supergiants as well. This means that none of the systems we
investigated has had a subdued recent star-formation rate.

The sample galaxies show, quite in general, a
large fraction of yellow supergiants, usually over 50\% and up to about
75\%. This is probably due to the poor resolution of
the galaxy images. The Hunter \& Gallagher (1985) spectroscopically investigated
dI sample has, on average, only a little over 20\% of AF stars and they 
find about 35\% of OB stars and about 45\% of G-M stars as the typical fraction 
resulting from their population synthesis analysis. Due to the small numbers of 
resolved stars, we do not feel that our data can contribute to the question of
the ratio of blue-to-red suergiants in galaxies, one of the major unsolved problems
of stellar astrophysics. The blue-to-red supergiant ratio of galaxies is observed to be
in the range from 0.4 in the SMC, to 3.6 in young Galactic clusters, 
and is a function of metallicity (see Langer \& Maeder 1995). 
Six of our sample galaxies have had their HII-region abundances measured and show
low oxygen abundances with respect to solar.
Keeping in mind that in our data, the number of 
red supergiants is expected to be affected by incomplete foreground-star subtraction, 
and that the blue supergiant counts will include main-sequence stars,
our result for the average blue-to-red supergiant ratio of the sample
galaxies is consistent with published ratios. Clearly, data of higher
spatial resolution are needed to contribute to this interesting question.

\subsubsection{Intermediate-age and old stars}

We cannot distinguish stars such as asymptotic giant branch stars or
red giants, which would allow us to access older
stellar populations based on the CMDs. However, we can make use of the integrated
color of the background light to learn about the older stars. 

What is the stellar content indicated by the underlying light distributions?
We presented in section 5 our findings of
the regular, elliptical morphology of the background light distributions
and the large extent of most sample galaxies in the R band, as well as their
redder colors towards larger radii. This provides
some indication for the presence of dynamically older stellar components. 

As a first step towards an interpretation of the background-light colors, 
we can make the assumption that the bright, resolved stars belong to a 
population that is young, and that the faint light of the unresolved stars 
originates from a population that is old.
Using the Johnson (1966) Tables, the colors of the underlying light 
distributions correspond typically to mid-F-type main sequence stars. Such 
stars have lifetimes of the order of several Gyr.
  
A more detailed interpretation of the color of the underlying light 
distributions, which accounts for its (presumed) low metallicity, can be 
carried out using the population synthesis models of Schmidt et al. (1995). 
To model the color of an old, metal poor population, Schmidt et al. use a 
combination-spectrum of several low-metallicity, Galactic globular 
clusters. The B-R color of their cluster template is 1.18$^m$. 
A red color of the background light with B-R$>$1$^m$ is observed in two of our
sample galaxies, UGC~6456 and UGC~8091, suggesting the presence of
a stellar population substratum with an age of several Gyr to a Hubble time.

In the remaining galaxies, the background light is bluer. This either indicates
that the underlying stellar populations that we detect have been made less 
than a few times 10$^9$~yrs ago, in some cases, even less than a few times
10$^8$~yrs ago, or we must explore the idea of a more complex star
formation history. The next simplest step that we can take in the 
interpretation of the background-light colors is to assume the background
light itself is composed of a mix of at least two stellar generations. 
To do so we compare the background-light colors with the colors 
predicted by Schmidt et al. from a combination of the Galactic globular 
cluster template with that of young clusters having a range of ages. 
B-R colors in the range of 0.8$^m$ to 1$^m$, as seen in several sample galaxies, 
(see Table 4) can easly be reproduced with
a combination of the oldest- and various younger-population templates. 
Both the ages of star-forming events, and their strengths need to
be considered when predicting galaxy colors.
Specifically, the model colors will depend on just how much light
(or mass) is contributed by a younger population to dilute the
light from the old population; young populations contributing just 0.1\% to
10\% in mass and with ages from a few 10 Myrs to 500 Myr can thus
produce the colors of UGC~7559, UGC~4459, UGC~8769, and UGC~5272~A. 
An alternative interpretation of this result is that these galaxies have
experienced a more-or-less constant star-formation rate over long
time scales.

The fairly blue colors, in the range of 0.3$^m$ to 0.5$^m$, which we observe 
for the background-light distributions of UGC~8320, UGC~8024, UGC~5272~B and UGC~5340,
can only be produced (in the Schmidt et al. models) with a large mass fraction 
(10\% or more) of young stars. In such cases, the underlying old population
remains undetectable.

\subsubsection{Star-formation Histories}

An attempt at interpreting the results on the resolved stellar content and
the background-light colors in terms of the star-formation histories of 
the sample galaxies leads to the following picture. It is highly likely that
all of the sample galaxies contain old stars, i.e, with ages of the order
of a Hubble time, but that due to different mass fractions of stars produced throughout
the histories of individual galaxies, the light from these faint, red stars is diluted by
light from bluer and younger stars to various degrees. In
UGC~6456 and UGC~8091 we observe red background-light colors indicative of
old stars, and we also see HII regions and resolve young stars.
After their initial star-forming events,
these galaxies may have experienced a long period of quiescence before the onset 
of a more recent episode of star formation. 
UGC~7559, UGC~4459, UGC~8760, and UGC~5272A show resolved, young stars
and, in addition, intermediate colors of their background light, indicating that
the light of the old stars is partially diluted by that from younger stars. 
These galaxies may have been making stars more continuously, such as in a 
series of several events, in their recent histories. 
UGC~8320, UGC~8024, UGC~5272~B, and UGC~5340, showing the bluest colors of
their background light and also resolved, young stars, may have been producing 
large mass fractions of stars in their recent histories.

\subsection{Environmental Effects}

As a result of section 6.1, we found that UGC~5272~A and UGC~5340 are closer
and UGC~6456 is more distant than previously thought. In order to re-asses 
their isolation (see section 2), we investigated the
distribution of galaxies in KK along the line-of-sight toward these galaxies.
There are no galaxy groups or massive galaxies at the new distances, with
which these sample galaxies could be associated. 

Apart from the isolation of our sample galaxies from massive neighbors,
we also have to investigate proximity to other low-mass galaxies, in particular
since dwarf-dwarf interactions have been considered as star-formation
triggers in dwarf galaxies. We note that UGC~8320 is actually located near UGC~8308 (Im~V),
UGC 8331 (Im IV) and UGC 8215 (Im V); UGC~8760 could form a group
of dwarfs with UGC~8833 (Im pec) und UGC~8651 (Im IV), and 
UGC 5272~A with UGC 5272~B and UGC~5340 (all sample galaxies).
This suggests that in our sample, only UGC~6456 is a truly isolated field galaxy.

Having re-discussed the relative isolation of our sample galaxies, we compared several
of their global properties with those of other galaxy samples. Let us note
at the onset that we are dealing with highly uncertain results in this section,
owing to our small sample size, the sometimes ill-known parameters being compared,
and the fact that comparison samples were compiled from the literature.

\subsubsection{Effects of Massive Neighbors}

Structural types of massive, luminous galaxies are well known to vary as a function
of their environmental galaxy density. The debate over whether this difference
is due to instrinsic conditions or environmental factors, such as galaxy-galaxy
interactions, currently seems to favor the latter. We ask whether there is
any effect of environment on the structural parameters of dIs. 
The structural parameters of the sample galaxies were previously
compared with those of Virgo Cluster dwarfs by Hopp 
(1994). He showed that the same correlation of scale length with B$_T$ holds 
for field and cluster galaxies. In other words, 
the stellar components of isolated galaxies show no obvious 
structural differences from the Virgo Cluster members at least
with respect to size. We compared the structural parameters
SB(0)$_B$ and $\alpha$$^{-1}$$_B$ of our sample galaxies (see Table 3) with
those derived by Vennik et al. (1996) for galaxies distributed over a
large range of absolute blue magnitudes and situated in groups, sheets
and voids. The structural parameters of our sample galaxies overlap
with, and form a smooth extension of, the low-luminosity end of the
structural-parameter distributions derived for those galaxies. In
terms of size and central surface brightness, our sample galaxies show
no difference from them. The conclusion is that apparently, some
structural parameters of dIs such as their stellar-disk sizes are
insensitive to the environment.

Another issue related to structural parameters of the sample galaxies
(but not their environment) is their classification. We noted repeatedly
that some of the dIs in our sample have a BCD designation in the
literature (UGC~5272~A, UGC~5340, UGC~6456) and we mentioned
the lack of a clear dI vs. BCD classification criterion. Recently,
Meurer (1998) suggested that a line of demarcation could be drawn on
the basis of blue central surface brightness, with objects showing
SB(0)$_B$ brighter than about 22$^m$/arcsec$^{2}$ being BCDs and
objects fainter than about 22$^m$/arcsec$^{2}$ being dIs. While in
our sample, UGC~5340, listed as a type Im/BCD in the catalog of Binggeli et al. (1990),
has SB(0)$_B$ close to 22$^m$/arcsec$^{2}$, UGC~5272~A and UGC~6456
are somewhat fainter. However, inspection of Meurer's Fig.~3 suggests that 
there is actually some overlap between the dI and BCD galaxy types in the 
21.5-23.5$^m$/arcsec$^{2}$ region, and our sample galaxies happen to
fall with their SB(0)$_B$ exactly into this overlapping region.

The global colors of dIs reflect their populations and recent star-formation histories.
As a group, Irregular galaxies are the bluest of the normal galaxies, with
average colors of U-B=-0.3, B-V=0.4 (Hunter \& Gallagher 1986). Note that
among the IGs there may be an effect of bluer colors occuring among
galaxies of low luminosity, i.e., the dIs.
Gallagher \& Hunter (1985, 1987) used broad-band, total galaxy colors to compare
the stellar populations and star-formation rates of a sample of Virgo Cluster
IGs with a sample of IGs in the field (note, these samples comprise
not just dIs, but contain more luminous Irregulars as well). 
They found that Cluster members are, on average,
redder, U-B$\simeq$-0.13, than galaxies in the field sample, U-B$\simeq$-0.33. 
In Figure 35, we compare the total U, B, V, R colors of our sample galaxies (cf. HS) 
with those of the samples studied by Gallagher \& Hunter (1986, 1987).
Our galaxies tend to be found in the blue portion of the distributions, with
an average U-B=-0.35. Our sample galaxies show colors
that resemble those of the dIs or BCDs, and of IGs
found in the field. Since our sample was chosen to comprise dIs in low-density
regions, the dependence of IG colors on both galaxy luminosity and 
environmental density affect the result -- are these galaxies blue because all
dIs have blue colors, or are they blue because they are located in the
field? Maybe these two issues are not independent of one another.
In the Local Group, for instance, the numbers increase and the ages of the dominant
populations of low-mass galaxies decrease, with increasing distance from the center
(van den Bergh 1994). We conclude that the blue total colors of our sample galaxies
indicate that vigorous star formation does occur in isolated dIs, i.e., in 
the absence of external triggering by massive galaxies.

In the previous section, various indicators for young and
old populations were employed in an attempt glean some limited insight into the
star-formation histories of the sample galaxies. All of the galaxies show
HII regions and resolve into blue stars, hence all of them are actively forming
stars at the present epoch. However, where the star-formation histories are
concerned, there appears to be some variety both in the amount and in
the time-scales over which stars have been forming. This is rather similar
to results found for the Local Group dIs.
Their star formation took place at different times and with different
intensities as well (Grebel 1997). It is possible that the data quality and 
our understanding of the CMDs, as well as our knowledge of the history of interactions,
is not yet sophisticated enough to enable us to discern the role of 
envionmental effects on star-formation histories of dIs. Perhaps
the stellar population differences between cluster and field dIs, 
if they exist, are subtle, otherwise we would classify the galaxy 
as something else, such as, e.g., a dSph.

In order to continue actively forming stars, a galaxy needs to have
available a reservoir of material from which to form stars.  
Huchtmeier et al. (1997) compared the M$_{HI}$/L$_B$ ratios  
(considered an indicator for the star-forming potential of a galaxy) for a
sample of dwarf galaxies in voids, with other dwarf galaxy samples,
including that of dwarfs in the sample of nearby galaxies (KK) and of
Virgo cluster dwarfs.  They find a tendency for isolated galaxies to
show larger values of M$_{HI}$/L$_B$ compared to the other samples,
and suggest that this may be due to a smaller chance for galaxy-galaxy
interactions for the void galaxies. We have taken the ratio of 
HI-mass-to-blue-luminosity for our galaxies from the compilation Schmidt 
\& Boller (1992). The value of the mean
M$_{HI}$/L$_B$ of our sample galaxies is 1.98 (in units of of M$\sun$
and L$\sun$) and the scatter is $\pm$ 1.94. Hence, our sample galaxies
possibly also show an elevated M$_{HI}$/L$_B$ ratio compared to samples
of dwarfs in denser environments. The conclusion is that the
environment of our sample galaxies is in principle favorable for
fueling continuing star formation in the future.

In spite of their
isolation, our sample galaxies exibit globally very blue colors and
have large HI masses for their blue luminosities, indicators for current
and potential future high levels of star formation.
With the caveats outlined at the beginning of the section, our results
indicate that interactions with massive galaxies are unimportant
as a trigger of star-formation in the sample galaxies. But, just how much
the star-formation histories of field vs. cluster dIs differ as a result
of different interaction histories remains unclear. It is suggested
that one difference between galaxies in high-density and low-density environments
might be their gas reservoirs.
This could result in isolated dIs evolving more slowly, although that will also
depend on how they recycle, acquire, or lose ISM.

\subsubsection{Effects of Low-Mass Neighbors}

In the absence of massive neighbors, what may trigger the star formation
in dwarfs? Among the possibilities discussed in the literature
are the hypothesis of self-propagating star-formation, interaction
with low-mass neighbors, or the accretion of HI gas until a critical
threshold is exceeded (see section 1).

Two pieces of information on our sample galaxies appear relevant.
First, HII regions and young associations can
be spread out over an entire galaxy. This may present a problem
for the self-propagating star-formation scheme. How and why are such widely
distributed regions of a galaxy all forming stars at the same time, when  
the self-propagating star-formation hypothesis rather envisions star formation 
to migrate with time from one active site of a galaxy to an adjacent site?

Second, several galaxies show outlying associations of blue stars or outlying HII 
regions (Table 4). An interesting question that arises is the following: 
when should we call an object and outlying HII region/association of a given dI 
and when a dI in its own right? This needs to be 
disentangled in the context of the hypothesis that dwarf-dwarf 
encounters trigger star formation in dIs. An educational example is provided by
UGC~5272~A,B. Hopp \& Schulte-Ladbeck (1991) called one outlying association
of stars UGC~5272~B and thought of it as a companion galaxy. Another
star cluster, however, was described as an outlying association rather
than a companion galaxy, due to its greater proximity to UGC~5272~A. 
Note that Taylor et al. (1993) comment that UGC~5272~A and B are
too close to one another to be separated on their HI maps. Another instructive
example is that of the BCD SBS~0335-052 which also exhibits an
``outlying HII region"; this region is demonstrated to be
included within the HI envelope of the galaxy (Lipovetsky et al. 1996).
Where should one draw the line between an outlying association and a dI
companion? And if the interaction with an HI cloud triggers star formation
in the dI, shouldn't it also trigger star-formation in the HI cloud it is
interacting with (but see Chengalur et al. 1995)? 

The presence of the outlying associations shows that star formation in dIs is 
taking place in regions of a galaxy that are quite far from other regions 
of the same galaxy which have previously been forming stars; again, this 
would seem to present a problem for the self-propagating star-formation hypothesis. 
If any one mechanism is to explain the tiggering of star formation
in isolated dIs, it will have to account for the range of morphologies
seen in the distribution of star-forming sites, as well as for the variety
in star-formation histories.

\section{CONCLUSIONS}

We presented and discussed the morphology and nature 
of resolved stars in 10 dIs which are generally more distant than galaxies
previously resolved into single stars. We also discussed the
character of the underlying, extended light distributions of the
unresolved stars of these galaxies.  

Whereas the shape of the underlying light distributions can be modeled by
ellipses, the distribution of the bright, resolved stars is frequently clumpy.
HII regions and associated blue stars are present in all of the sample
galaxies, sometimes scattered across the 
entire face of the parent galaxy, sometimes clustered in distinct regions only.

Several cases of HII regions and young stellar associations were 
pointed out that are located outside of the main body of resolved
stars of the parent galaxy, but not outside of the Holmberg radius. 
The question of where to draw the dividing line
between an ``outlying association" and a ``companion dI" was raised.
This problem has obvious implications regarding the hypothesis that 
encounters with other dwarf galaxies or HI clouds trigger star formation in dIs. 
On the other hand, the self-propagating star-formation hypothesis implies that
star-forming regions in dIs should be found within some proximity 
of one another, rather than isolated. Why then do we see single aggregates of
young stars in remote regions of dIs? It should be interesting to compare the stellar
content of such outlying associations with that of more centrally located
star-forming regions.

The general picture that emerged is that while all sample galaxies 
are presently experiencing star formation, their histories
were quite diverse, with differences in star-forming intensity and duration 
from galaxy to galaxy. 
A similar result has been derived for the dIs of the Local Group. With
current knowledge about the star-formation histories of dIs it is not
possible to distinguish any differences in evolution between
dIs in high- and low-density environments.

\section{Acknowledgments}

Support for this work was provided in part by NASA through grant number 
AR-06404.01-95A from the Space Telescope Science Institute, which is operated by 
the Association of Universities for Research in Astronomy, Incorporated, 
under NASA contract NAS5-26555. U.H. acknowledges support from the 
Sonderforschungsbereich 375 of the Deutsche Forschungsgemeinschaft.
Mr. Ashwin Vaidya helped with the data reduction and generated some of the
Figures. We thank W. Freedman for providing us with her IC~1613 data.
This research made use of the SIMBAD database, which is
operated by CDS, Strasbourg, France. Notice that
the Tables containing the single-star photometry discussed in this
paper are available through the CDS. We thank the anonymous referee for
the very detailed and explicit comments on this paper.

{}

\clearpage

{\bf FIGURE CAPTIONS}

\figcaption[sl_fig1.ps]{Illustrative template CMDs constructed from data 
in Schmidt-Kaler (1965).
The range of spectral types is from O9.5 to M4. Between B0 and M0, data
were considered in intervals of 0.5 subtypes. The luminosity classes
used were Ia, III, and V. The template CMDs are plotted to the same
scale as the CMDs of the sample galaxies. Overlaying the templates
onto the sample CMDs serves as an indicator for the envelope of
brightest stars in the galaxies.}

\figcaption[sl_fig2.ps]{The brightness-coded position of all sources
detected in the B-filter. The areas on the chip assumed to include
stars in the galaxy vs. stars in the foreground are separated by a
dashed line.}

\figcaption[sl_fig3.ps]{The color-coded position of all sources
detected in the B- and R-filters. Three symbol sizes are used. In order of
decreasing size the symbols indicate B-R$<$0.0, 0.0$\leq$B-R$<$1.0,
and B-R$\geq$1.0. This provides a rough separation of the stellar types
into OB, AF, and GKM stars.  }

\figcaption[sl_fig4.ps]{The panels on the left-hand side show
B vs. B-R CMDs, whereas the panels on the right-hand side show R vs. B-R
CMDs. The two top panels include all stellar sources detected in 
the B- and R-filters. }

\figcaption[sl_fig5.ps]{The I-filter image of UGC~4459 is shown on the 
bottom, the H$_{\alpha}$-I filter image is shown on the top.}

\figcaption[sl_fig6.ps]{The brightness-coded position of all sources
detected in the B-filter. The areas on the chip assumed to include
stars in the galaxy vs. stars in the foreground are separated by a
dashed line.}

\figcaption[sl_fig7.ps]{The color-coded position of all sources
detected in the B- and R-filters. Three symbol sizes are used. In order of
decreasing size the symbols indicate B-R$<$0.0, 0.0$\leq$B-R$<$1.0,
and B-R$\geq$1.0. This provides a rough separation of the stellar types
into OB, AF, and GKM stars.  }

\figcaption[sl_fig8.ps]{The panels on the left-hand side show
B vs. B-R CMDs, whereas the panels on the right-hand side show R vs. B-R
CMDs. The two top panels include all stellar sources detected in 
the B- and R-filters. }

\figcaption[sl_fig9.ps]{The I-filter image of UGC~7559 is shown on the 
bottom, the H$_{\alpha}$-I filter image is shown on the top.}

\figcaption[sl_fig10.ps]{The brightness-coded position of all sources
detected in the B-filter. The areas on the chip assumed to include
stars in the galaxy vs. stars in the foreground are separated by a
dashed line.}

\figcaption[sl_fig11.ps]{The color-coded position of all sources
detected in the B- and R-filters. Three symbol sizes are used. In order of
decreasing size the symbols indicate B-R$<$0.0, 0.0$\leq$B-R$<$1.0,
and B-R$\geq$1.0. This provides a rough separation of the stellar types
into OB, AF, and GKM stars.  }

\figcaption[sl_fig12.ps]{The panels on the left-hand side show
B vs. B-R CMDs, whereas the panels on the right-hand side show R vs. B-R
CMDs. The two top panels include all stellar sources detected in 
the B- and R-filters. }

\figcaption[sl_fig13.ps]{The I-filter image of UGC~8024 is shown on the 
bottom, the H$_{\alpha}$-I filter image is shown on the top.}

\figcaption[sl_fig14.ps]{The brightness-coded position of all sources
detected in the B-filter. The areas on the chip assumed to include
stars in the galaxy vs. stars in the foreground are separated by a
dashed line.}

\figcaption[sl_fig15.ps]{The color-coded position of all sources
detected in the B- and R-filters. Three symbol sizes are used. In order of
decreasing size the symbols indicate B-R$<$0.0, 0.0$\leq$B-R$<$1.0,
and B-R$\geq$1.0. This provides a rough separation of the stellar types
into OB, AF, and GKM stars. }

\figcaption[sl_fig16.ps]{The panels on the left-hand side show
B vs. B-R CMDs, whereas the panels on the right-hand side show R vs. B-R
CMDs. The two top panels include all stellar sources detected in 
the B- and R-filters. }

\figcaption[sl_fig17.ps]{The brightness-coded position of all sources
detected in the B-filter. The areas on the chip assumed to include
stars in the galaxy vs. stars in the foreground are separated by a
dashed line.}

\figcaption[sl_fig18.ps]{The color-coded position of all sources
detected in the B- and R-filters. Three symbol sizes are used. In order of
decreasing size the symbols indicate B-R$<$0.0, 0.0$\leq$B-R$<$1.0,
and B-R$\geq$1.0. This provides a rough separation of the stellar types
into OB, AF, and GKM stars. }

\figcaption[sl_fig19.ps]{The panels on the left-hand side show
B vs. B-R CMDs, whereas the panels on the right-hand side show R vs. B-R
CMDs. The two top panels include all stellar sources detected in 
the B- and R-filters. }

\figcaption[sl_fig20.ps]{The brightness-coded position of all sources
detected in the B-filter. The areas on the chip assumed to include
stars in the galaxy vs. stars in the foreground are separated by a
dashed line.}

\figcaption[sl_fig21.ps]{The color-coded position of all sources
detected in the B- and R-filters. Three symbol sizes are used. In order of
decreasing size the symbols indicate B-R$<$0.0, 0.0$\leq$B-R$<$1.0,
and B-R$\geq$1.0. This provides a rough separation of the stellar types
into OB, AF, and GKM stars.  }

\figcaption[sl_fig22.ps]{The panels on the left-hand side show
B vs. B-R CMDs, whereas the panels on the right-hand side show R vs. B-R
CMDs. The two top panels include all stellar sources detected in 
the B- and R-filters. }

\figcaption[sl_fig23.ps]{The I-filter image of UGC~5340 is shown on the 
bottom, the H$_{\alpha}$-I filter image is shown on the top.}

\figcaption[sl_fig24.ps]{The brightness-coded position of all sources
detected in the B-filter. The areas on the chip assumed to include
stars in the galaxy vs. stars in the foreground are separated by a
dashed line.}

\figcaption[sl_fig25.ps]{The color-coded position of all sources
detected in the B- and R-filters. Three symbol sizes are used. In order of
decreasing size the symbols indicate B-R$<$0.0, 0.0$\leq$B-R$<$1.0,
and B-R$\geq$1.0. This provides a rough separation of the stellar types
into OB, AF, and GKM stars.  }

\figcaption[sl_fig26.ps]{The panels on the left-hand side show
B vs. B-R CMDs, whereas the panels on the right-hand side show R vs. B-R
CMDs. The two top panels include all stellar sources detected in 
the B- and R-filters. }

\figcaption[sl_fig27.ps]{H$_{\alpha}$ image of UGC~6456.}

\figcaption[sl_fig28.ps]{The brightness-coded position of all sources
detected in the B-filter. The areas on the chip assumed to include
stars in the galaxy vs. stars in the foreground are separated by a
dashed line.}

\figcaption[sl_fig29.ps]{The color-coded position of all sources
detected in the B- and R-filters. Three symbol sizes are used. In order of
decreasing size the symbols indicate B-R$<$0.0, 0.0$\leq$B-R$<$1.0,
and B-R$\geq$1.0. This provides a rough separation of the stellar types
into OB, AF, and GKM stars.  }

\figcaption[sl_fig30.ps]{The panels on the left-hand side show
B vs. B-R CMDs, whereas the panels on the right-hand side show R vs. B-R
CMDs. The two top panels include all stellar sources detected in 
the B- and R-filters. }

\figcaption[sl_fig31.ps]{H$_{\alpha}$ image of UGC~8320. }

\figcaption[sl_fig32.ps]{The brightness-coded position of all sources
detected in the B-filter. The areas on the chip assumed to include
stars in the galaxy vs. stars in the foreground are separated by a
dashed line.}

\figcaption[sl_fig33.ps]{The color-coded position of all sources
detected in the B- and R-filters. Three symbol sizes are used. In order of
decreasing size the symbols indicate B-R$<$0.0, 0.0$\leq$B-R$<$1.0,
and B-R$\geq$1.0. This provides a rough separation of the stellar types
into OB, AF, and GKM stars. }

\figcaption[sl_fig34.ps]{The panels on the left-hand side show
B vs. B-R CMDs, whereas the panels on the right-hand side show R vs. B-R
CMDs. The two top panels include all stellar sources detected in 
the B- and R-filters. }

\figcaption[sl_fig35.ps] {Two-color diagrams comparing the total,
intrinsic galaxy colors for our sample galaxies (filled squares) with
the colors of IGs from Gallagher \& Hunter (1986) (small filled dots),
dIs in the field (filled triangles) and Virgo cluster IGs (open triangles)
from Gallagher \& Hunter (1987). 
The colors of the sample galaxies are noticably blue.}

\end{document}